\documentclass[onecolumn,11pt]{IEEEtran}
\usepackage{comment}
\usepackage{graphicx}
\usepackage{amsfonts,amsmath}

\def\E{{\sf E}}
\def\P{{\sf P}}

\def\eg{{\em e.g.},~}
\def\ie{{\em i.e.},~}
\def\cf{{\em cf.},~}
\newcommand{\beqa}{\begin{eqnarray*}}
\newcommand{\eeqa}{\end{eqnarray*}}
\newcommand{\be}{\begin{eqnarray}}
\newcommand{\ee}{\end{eqnarray}}

\begin{document}

\title{Numerical Evaluation of Cloud-Side Shuffling Defenses against DDoS 
Attacks on Proxied Multiserver Systems\thanks{This research supported in part by DARPA XD3 grant, 
Cisco Systems URP gift, and EC2 and GCE credit gifts.}}

\author{
\begin{tabular}{cc}
Yuquan Shan, George Kesidis & 
Daniel Fleck, Angelos Stavrou\\
School of EECS & CS Dept \\
Pennsylvania State University  & George Mason University\\
 University Park, PA, USA & Fairfax, VA, USA\\
\{gik2,yxs182\}@psu.edu  & 
\{dfleck,astavrou\}@gmu.edu
\end{tabular}
}

% The default list of authors is too long for headers}
%\renewcommand{\shortauthors}{Y. Shan et al.}

\maketitle

\begin{abstract}
We consider a cloud based multiserver system, that may
be cloud based, consisting of a set of replica
application servers behind a set of proxy (indirection) servers
which interact directly with clients over the Internet.
We address cloud-side proactive and reactive defenses
to combat DDoS attacks that may target this system.
DDoS attacks are endemic with some notable attacks
occurring just this past fall. Volumetric attacks may
target proxies while ``low volume" attacks may target
replicas.
After reviewing existing and proposed defenses, such  as
changing proxy IP addresses (a  ``moving target" technique
to combat the reconnaissance phase of the botnet) 
and fission of overloaded servers,  we focus on  evaluation of
defenses based on shuffling client-to-server assignments
that can be both proactive and reactive to a DDoS attack.
Our evaluations are based on a binomial distribution model
that well agrees with simulations and preliminary
experiments on a prototype that is also described.
\end{abstract}


%\acmBadgeR{artifacts_available}

%\maketitle

%\input{../fission/intro}
\section{Introduction}

\begin{figure}[h]
\begin{center}
\includegraphics[width=\columnwidth]{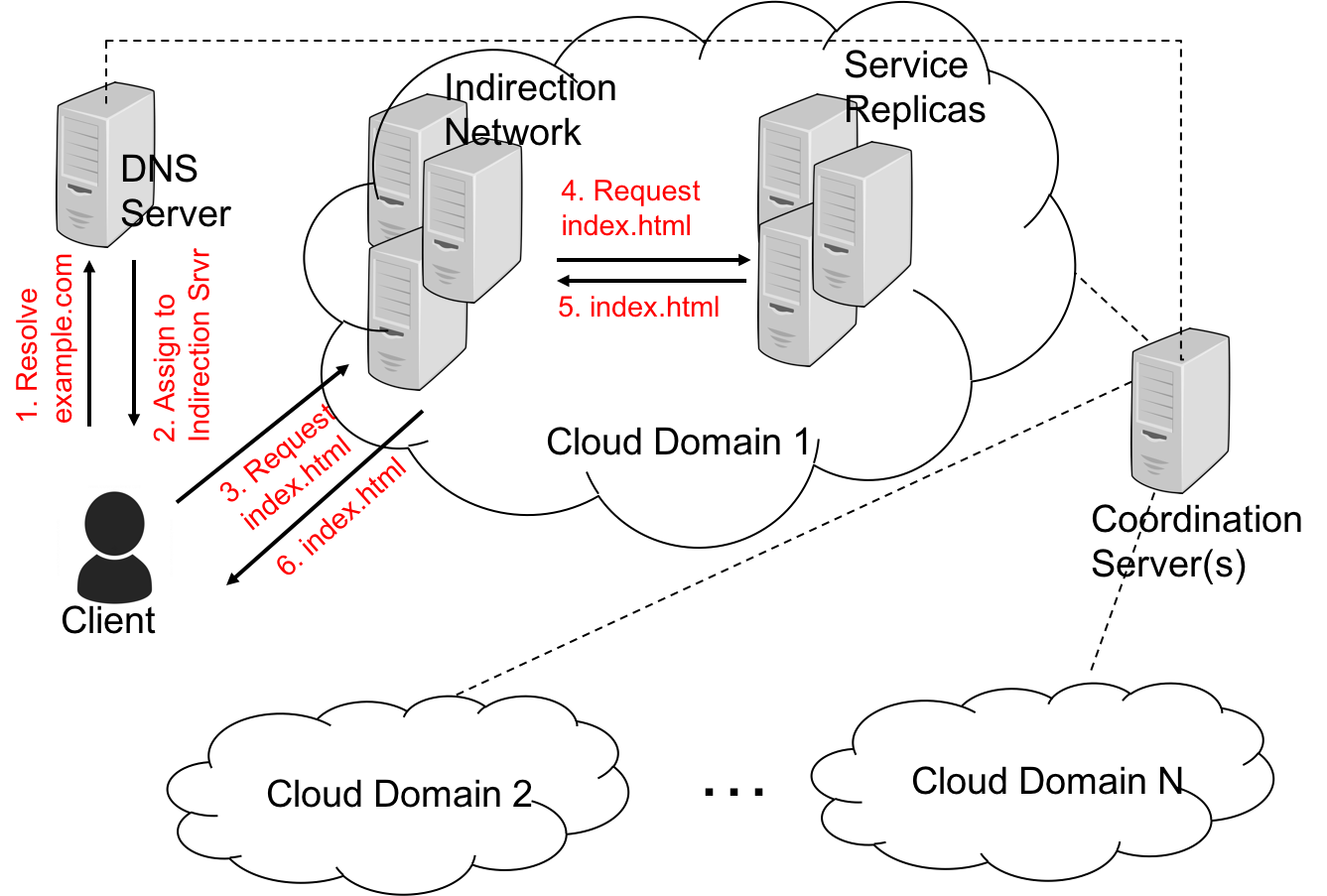}
\caption{Overview of system architecture}\label{fig:overview-sys-arch}
\end{center}
\end{figure}

Consider a generic multiserver virtual system of clients, indirection/proxy servers,
and replica worker/application servers (replicas), where
the servers are controlled
either by  a central coordination server
%(\eg via round-robin DNS)
or in a distributed fashion  by the proxies themselves
 \cite{Stavrou13,Stavrou14,Stavrou16,Jajodia16},
see Figure \ref{fig:overview-sys-arch}.
%All of the aforementioned coordination, proxy or
%application/replica ``servers'' are implemented in virtual
%machines or containers inside of {\em physical} servers of
%one or more data-centers.
The clients reach the proxies via the public commodity Internet.
The proxies and replicas could be implemented
in Virtual Machines (VMs) or lightweight containers (instances)
on physical servers
of one or more datacenters of the public cloud.
The proxies in turn
assign clients to replicas to fulfill their
requests.
The clients never know the IP addresses
of the replicas, only those of the proxies which they determine
by DNS, where the coordination server manages the DNS records of the proxies.
That is, the proxies segment the sessions between clients
and replicas.
For example, load-balanced content distribution  networks
%or general purpose ``derivative clouds'' 
or session establishment  systems
can be mounted on such systems.
%CDNs Akamai or Netflix? DNS system? SIP servers 

%``Timeline of system events''  figure too? from PI meeting slidedeck
% ../slides/timeline-events.png

Generally, the threat of Distributed Denial of Service 
(DDoS) attacks (\eg \cite{dostax,Mirkovic05,Kesidis06})
continue to be a concern because the fundamental lack of economic 
incentives
to secure attack-participating bots\footnote{\eg
well known techniques
of specification-based 
behavioral anomaly detection and prevention for IoT devices, 
\eg \cite{Sanders11,Rowe16}} 
and/or deploy egress filtering in their access network. 
% incl. wireless router
% \eg \cite{Krebs16}.
Indeed, two very significant botnet based DDoS attacks were
witnessed just this past Fall \cite{DDoS-Dyn-102116,Krebs16}.

In Section \ref{sec:defense-overview},
for a generic proxied multi-server system,
%consisting
%of a coordination server, proxy servers and application server replicas,
we overview proposed {\em cloud-side} defenses to combat
\begin{itemize}
\item the reconnaissance phase of a botnet to identify the current
IP addresses of the proxies,
\item high-volume floods targeting the proxies
(\eg recently \cite{DDoS-Dyn-102116,Krebs16}), and
\item low-volume attacks targeting the replicas
(\eg \cite{Slowloris} 
%like a syn flood 
and more recently \cite{BlackNurse}).
%like ping of death
\end{itemize}
These cloud-side defenses include 
periodically changing IP addresses of the proxies
(to proactively address botnet reconnaissance) \cite{Stirling-TR},
fission of overloaded servers \cite{fission-TR}, and
dynamically shuffling
client-to-server assignments
\cite{Stavrou13,Stavrou14,Stirling-TR} 
(requiring migration of client state \cite{clark2005live}).

The focus of this paper is shuffling based defenses.
In Section \ref{sec:def}, we give some preliminary definitions
of the binomial model to be used in subsequent system models.
In Section \ref{sec:shuffling}, we assess client-to-server shuffling,
including for proactive load balancing and  leveraging 
the clients' history of assignment to overloaded servers (client
reputations).
How shuffling can be used to quarantine heavy-hitter/attacker clients
is evaluated in Section \ref{sec:shuffling-quarantine}.
We conclude with a summary in Section \ref{sec:summary}.

\section{Summary overview of considered threats and proposed cloud-based defenses}\label{sec:defense-overview}

DDoS defense is closely related to load-balancing operations.
Among available servers,
load balancing may {\em distribute}  the ``heavy 
hitter'' clients (including assigning a dedicated server)
as well as evenly distribute more numerous clients with light workloads.
If heavy hitters are violating some terms of use or SLA, they may be 
individually deemed abnormal and  throttled or consolidated (quarantined). 
An individual client's workload may naturally vary over time and 
a few heavy hitter clients at a given point in time may be considered normal
and may not result in significant performance degradation of the servers
%(unless the heavy hitters happen to be collocated).
under load balancing.

When there are simultaneously significantly more sustained heavy hitters or 
significantly more light-workload clients than normal,  so that
the performance of a {\em sufficiently large number of servers  is 
significantly degraded (overloaded servers)},
then a DDoS attack or flash crowd may be occurring.
In this case, it is desirable
that deemed heavy hitters be terminated, throttled or
consolidated in quarantine (effectively underserved),
\ie heavy hitter clients would then be considered attackers.
Note that load-balancing operations alone
may address many weaker DDoS attacks - such 
``missed detections" are not as important because 
the resulting performance degradation of the servers is not significant.
Clients deemed attackers may be challenged, \eg
Javascript puzzles, that likely cannot be solved by bots running
on IoT devices deployed with limited protocol suites  - in this
way, a DDoS attack and a flash crowd can be discriminated.

In the following, we will describe mechanisms that can be used:
\begin{itemize}
\item {\em proactively} as a load balancer
before a DDoS attack is detected to deal with
the problem overloaded servers by distributing
heavy-hitter clients 
among all available servers (possibly by identifying
heavy-hitter clients in the first place); or 
\item {\em reactively} after a DDoS attack is detected,
\eg to quarantine deemed attacking clients.
\end{itemize}
In some cases, the same techniques (\eg shuffling
client-to-server assignments) can be used both proactively
and reactively.

\subsection{General assumptions of attack and defense on servers}

To reduce costs, multiple client end-users are assigned to each
server and individual clients are not ``containerized'' within each server
(Virtual Machine).
A key assumption in the following is that an overloaded (attacked) server 
cannot determine which of its assigned client(s) are responsible.
This said, it can be reliably determined whether a server itself
is overloaded, \cf Section  \ref{sec:overhead}. 

In some defense frameworks, new clients are never added 
to servers (either proxies or replicas)
deemed overloaded. %, unlike \cite{Jajodia16}.
%\cite{Jajodia16} involves
%no collusion, instant migration, new clients go where old clients are
%(in particular clients affected by attacks).
``Liberated'' nominal (benign) clients may be consolidated into fewer
servers to reduce the number of virtual machines or
containers  (instances) used by the overall
system, particularly when the threat model is such that a {\em single}
attacker can detectably affect a server (as assumed for application
server replicas). However, as discussed below, individual proxy servers
subjected to flooding attacks may be able to tolerate more than one 
attacker  ($A>1$) - so consolidating proxy servers deemed {\em not} under 
attack may result in servers that {\em are} subsequently detected under attack.
%Finally, note that a proxy subjected to a DDoS flood may be able to
%tolerate more than one attacker,  \ie 
%attack detection based on sufficient performance degradation
%may not occur unless a server is subjected
%to $A>1$ attackers.

Some attacking clients are simple bots that cannot
cope with reassignment to a new server.  For such bots,
continual shuffling of clients to servers may be a sufficient
defense.  
We focus  herein on other malware 
that is more resilient and can continue its attack even after
 it is transferred to another server (or after the server modifies
its IP address), 
%\ie ``following bots," 
and on nominal clients that
can also be redirected to new servers.

%Both proactive moving-target and reactive
%quarantine defenses are considered herein.
 
\subsection{Proactive moving-target defense against 
botnet reconnaissance phase}
\label{sec:proactive-mtd}

The botmaster needs to gather the {\em current} IP addresses 
of as many proxies 
as possible before launching its DDoS attack. 
%During its reconnaissance
%phase, bots may need to engage in sessions of ``normal'' length and activity 
%to avoid detection of reconnaissance (or to      avoid blacklisting).  
Suppose the coordination server attempts to protect against botnet 
reconnaissance by periodically changing (via DNS) proxy IP addresses.  
%, including the use of round-robin records.
%In one approach, if a proxy is designated by the coordination
%server to have its IP address
%changed at time some time $t$ in the future, it receives no
%new clients in the time period $[s,t]$ where $t-s$ may be taken 
%as longer than a nominal client's session duration. 
Proxies may change their IP addresses individually (even at random) 
or  collectively at the same time.
In the latter case, {\em all} proxies known by the botnet 
become stale at once.
Clients with active sessions during a proxy's IP address change may
be interrupted and the proxy may or may not assist clients in re-establishing
their interrupted sessions.
Some bots may not be able to re-establish sessions and need to 
launch new ones, while nominal clients should be able 
reestablish sessions with only minor delay. 
In the Appendix, we describe  a simple model of this moving-target defense.

Each proxied multiserver system is a tenant of a datacenter.
Suppose that, as a service, the datacenter makes available a large pool of
external IPv4 IP addresses for proactive moving-target defense.
Tenants wishing to engage in moving-target defense would dynamically
share the entire pool with other tenants.
This would be more effective and less costly than the
individual tenants managing their own sets of addresses that would
need to be much larger in number
than their proxies, or periodically asking
the datacenter for fresh addresses.
Naturally, under IPv6, a vast number of addresses would be available
so that each tenant could cost-effectively mount such moving-target defense
on their own.

%Simple models (\eg birth-death Markov chains modeling the current 
%number of proxies known  to the botnet or \cite{Stirling-TR}) 
%can capture how the performance of this system depends on the
%frequency at which proxy addresses are changed and the frequency
%at which new proxies can be identified by the botnet. 

\subsection{Fission of application replicas servers against low-volume attacks targeting them}\label{sec:fission0}

Low-volume threats 
are such that just one attacker can take down a replica application server.
The idea of a $m$-ary fission defense \cite{fission-TR} is that $m-1$ containers
are spun-up  for each container housing a server deemed overloaded
(and this typically on the same physical server).
The clients assigned to the overloaded server are equally divided among the
$m$ resulting containers, and overload (and DDoS 
attack) detection is repeated. Note that the fissioned
containers may or may not be on the same server.
%Fission may be used in concert
%with migration of clients to different servers and so may be part of 
%a proactive load-balancing system.

Even if they remain on the same server,
each fissioned containers is basically a group of processes within the
virtual machine running the server, and namespaces and cgroups
can be used to isolate different containers from each other.
For example, consider a server with four clients including a
BlackNurse client \cite{BlackNurse} which consumes CPU resources. By fissioning
into two containers hosting two clients each, the  container
hosting two nominal clients will have CPU resources that
are protected from the one 
hosting the attacker, \ie two clients were ``liberated" by fissioning.
If the attacker is instead a Slowloris \cite{Slowloris} client consuming
memory for half-open HTTP connections, this same benefit
of binary fissioning would result.

Containers housing only nominal clients (\ie containers deemed not 
overloaded) may be consolidated to reduce
the number of containers in play.
Fission and overload-detection are repeated until the attacking clients are
sufficiently quarantined (spinning up new containers can
be performed in parallel with detection).

Recall that the botnet is limited in that clients are not aware
of the identities of the application replica servers. So, they
are not aware whether plural bots share a replica.
%Again, a basic assumption of fission defense is that always
%individually containerizing
%every client within each server
%is too costly from a performance (and detection) point-of-view.

\subsection{Existing defenses against high-volume attacks targeting proxies}\label{sec:existing-defenses}

Classical high-volume attacks include TCP SYN  floods and
reflector attacks. 
Load balancing and traffic filtering 
techniques have been deployed to deal with attacks attempting 
to exhaust network bandwidth.  These techniques can be at a third-party 
system, like Akamai's Prolexic Proxy, located elsewhere in the Internet,
or by the public-cloud providers themselves.

\paragraph{Scaling the system volume with load balancing} 
In GCE, the frontend infrastructure can automatically scale to absorb certain types of attacks (\eg SYN floods), and if the user has provisioned a sufficient number of instances, the autoscaler can ramp up the backend servers to accommodate the traffic spikes of the attack \cite{gce-practice}. In AWS, Amazon CloudWatch alarms are used to initiate the autoscaling of the size of a user's Amazon EC2 fleet in response to user-defined events \cite{aws-practice}.

\paragraph{Detecting and filtering excess traffic} 
There is a non-user-configurable DDoS protection layer in an Azure user's virtual network. If a proxy server with an  Internet-facing IP address was overloaded by a DDoS attack, the DDoS protection layer would attempt to
detect the sources of attack\footnote{\eg by a challenge-response mechanism designed to
defeat a bot, or by SYN cookies (randomized
initial TCP sequence numbers).} and scrub the offending traffic \cite{azure-practice}. In GCE, anti-spoofing protection is provided for a user's virtual network by default \cite{gce-practice}. In AWS, Elastic Load Balancing (ELB) accepts only well-formed TCP connections to protect users from SYN floods or UDP reflection attacks; Amazon CloudFront can automatically terminate connections from ``slow reading/writing" attackers (\eg Slowloris) \cite{aws-practice}. Users can also define firewall rules for their virtual network to block the source IP addresses that are engaging in the attack or restrict the access to some ports \cite{aws-practice, gce-practice, azure-practice}. 

\paragraph{Changing proxy addresses}
Note that the moving-target defense of Section \ref{sec:proactive-mtd}
 would protect newly spun-up proxies in response to a DDoS attack, and
continue to protect the original proxies post-attack as well.

\subsection{Summary of additional proposed defenses against high-volume attacks 
based on shuffling client-to-server assignments}\label{sec:shuffling-overview}

   Suppose all clients using proxies 
%(of our proxied multiserver system)
that are overloaded could be periodically 
reassigned to (shuffled among) the attacked proxies at random.
By chance,
some formerly overloaded proxies are  assigned  only nominal clients and
those clients have thus been effectively ``liberated." 
As with the fission defense, 
overload/attack detection and shuffling among overloaded/attacked servers
is repeated until attacking/heavy-hitter clients are sufficiently quarantined.

Shuffling client-to-server assignments
can also be used on replica application servers
to protect against low-volume DDoS attacks targeting them.
Shuffling can be used in concert with a client reputation system
based on the history of client involvement with overloaded
servers.
Here, {\em all} client-to-server assignments are shuffled, even
those of servers not deemed overloaded. Clients who are
repeatedly assigned to servers thereafter deemed overloaded have decreasing
reputation relative to those that are often assigned to servers
deemed not overloaded. After several shuffles, clients with sufficiently
low reputation can be deemed attackers (respectively, heavy-hitters)
and quarantined (distributed/load-balanced among the active servers, 
respectively).
If {\em all} servers are initially under attack then:
\begin{enumerate}
\item sequentially sequester/queue a fraction of all clients to an idle server
until not all servers under attack;
\item employ shuffling/detection with reputation to quarantine some of the
remaining clients; 
\item introduce some of the sequestered clients into service; and repeat 1-3.
\end{enumerate}
Note that attack detection may not be perfect so an operator may not
want to identify as heavy-hitters/attackers just  the
clients with the lowest reputations.

\begin{comment}
%see ../reputations/reputation.tex
We are currently studying the following 
two-phase detection/quarantine reputation framework.
Upon attack detection,
suppose \underline{all} servers are shuffled even those
not detected as under attack, \ie \underline{initially} the aim is
not quarantine of the attackers.
After each shuffle, attack detection is performed on each server.
After each attack detection, the reputation of all clients
in attacked servers is reduced relative to those in not-attacked servers.
After a certain number of shuffle/detection steps, 
all clients with sufficiently low reputation are quarantined.
Note  that detection may not be perfect,  so 
may want to quarantine not just
the clients with {\em lowest} reputation.
If {\em all} servers are initially under attack then
\begin{itemize}
\item sequentially sequester/queue a fraction of all clients to an idle server
until not all servers under attack,
\item employ shuffling/detection with reputation to quarantine some of the
remaining clients, 
\item introduce some of the sequestered clients into service and repeat.
\end{itemize}

%Might instead assign a single reputation to clients with common address prefixes (as reputations are commercially used for email spam)
\end{comment}

%\subsection{Proactive shuffling client-to-server assignments}

Finally, shuffling client-to-server assignments can occur {\em proactively}, \ie before a DDoS attack is detected according to sufficiently
large number of servers deemed overloaded. 
Again, shuffling can be used as part of a load-balancing mechanism to {\em distribute}, rather than quarantine, deemed heavy-hitter 
(low reputation) clients. 
By continually distributing clients that are causing servers to slow
down, one is increasing the work-factor of a potential attacker. 
And again, once a DDoS attack is detected,
the policy of distributing heavy-hitter clients can 
be switched to quarantine.

\subsection{Overhead associated with defense}\label{sec:overhead}

Again, a key assumption is that a server is overloaded
cannot determine which of its assigned client(s) are responsible.
This said, it can be reliably determined whether a server
is overloaded either through the use of:
\begin{itemize}
\item probing ``canary'' clients
that can determine when response times of their mock workloads
have grown too high, 
\item heartbeat signals between proxy/replica servers and coordination server,
\item host-based (HIDS) security checks by  the OS of a VM 
managing a number of containers\footnote{or
by the  hypervisor of a physical server managing a number of VMs, \ie
Security-as-a-Service}
in which proxies or replicas operate.
This could be part of general server and client
diagnostic performance checks that are continually run by
many public-cloud providers (\eg AWS CloudWatch).
\end{itemize}
Response to volumetric attacks  may
require employing different physical machines if the downlink (ingress) 
network I/O is partitioned among its VMs/containers,
recall Section \ref{sec:existing-defenses}.
%Response may include hot-spare servers that are
%kept on standby, \ie idling hot spares. To fairly account
%for the use of hot spares,
%%servers  in shuffling response to a volumetric attack, 
%one ought to compare performance with $M$ servers and no hot spares
%against the use of $M-H$ servers with $H$ hot spares, \eg
%\cite{Stirling-TR}.

Note that if fission is used for high-volume attacks targeting
proxies, more specifically targeting (external) network I/O, then the new
containers may need to reside on separate physical machines for
subsequent attack detection. In this case, fission defense would be
much more complex and costly. 
Fission defenses for application servers might only
involve spinning up containers
or dynamically ``containerizing'' clients within their existing
virtual machines and not migrating them to application servers on
other physical machines.

Of course, the overhead of defense also
includes the IT resources required for the  defense
itself (\eg network I/O for stateful clients, disk I/O)
and service outages 
experienced especially by stateful nominal clients.
Note that, generally, clients have little state in proxies so shuffling
clients among proxies would require relatively little overhead compared
to migrating stateful clients among replica application servers
\cite{clark2005live,hines2009post} residing on different physical machines.
For example, a SIP server could have little state (once established,
the media between clients is exchanged directly without the SIP server's
involvement); however, some SIP servers monitor their
active sessions, \eg for billing purposes.   For another example,
clients of one streaming-media server could easily be migrated to
another that is mounted on a physical machine that has access to a (possibly
different) disk storing the same media being streamed;
in this case,
server migration may include suitably priming the cache of the
new server.

%\input{../fission/experimental-setup}
% !TeX root = hotcloud17.tex
\section{Developing experimental set-up}\label{sec:fleck}

There is substantial prior work on  DDoS attack
experimentation 
\cite{pft,tridentcom-jelena,milcom,qop,expcs} spanning
simulation, emulation, benchmarks and metrics. We are developing 
our defense system prototype on the Amazon AWS platform.
The system uses a combination of Linux Virtual Machines (VMs) and Docker containers to 
implement and test the defense approach.

\subsection{Defense Experimental Setup}

The defense consists of a layer of indirection proxies which are used to facilitate shuffling of 
clients and also serve to hide the application replica layer. The indirection proxies are each launched
from an indirection manager VM. A single VM will run an indirection manager and many indirection server 
Docker containers. The application replica layer is similar. A single VM will run a replica manager process
which then launches multiple replica Docker containers within the VM.  Each replica container executes a 
small management process and the protected service (e.g. a webserver).

To scale the system, any number of VMs can be launched and in turn add replica containers and indirection
proxy containers into the defense. 

The overall defense is controlled by another VM running a coordination server. This server is used to assign
clients to both replica and indirection containers. Additionally, state information about each container and VM 
comes to the coordination server which can bring new servers online and remove others as needed.

To facilitate testing in the prototype, a DNS server is also used which allows the coordination server to register
DNS names for each of the defense servers. The initial contact from the client comes to a forwarding
server which redirects the client to the initial proxy location based on input from the coordination server.  To fully simulate a 
typical $n$-tier architecture we have also added a back-end MySQL database VM accessible by the web tier on the 
replica containers. The full architecture is shown in Figure \ref{fig:physical-arch} below.

\begin{figure}[h]
\begin{center}
\includegraphics[width=\columnwidth]{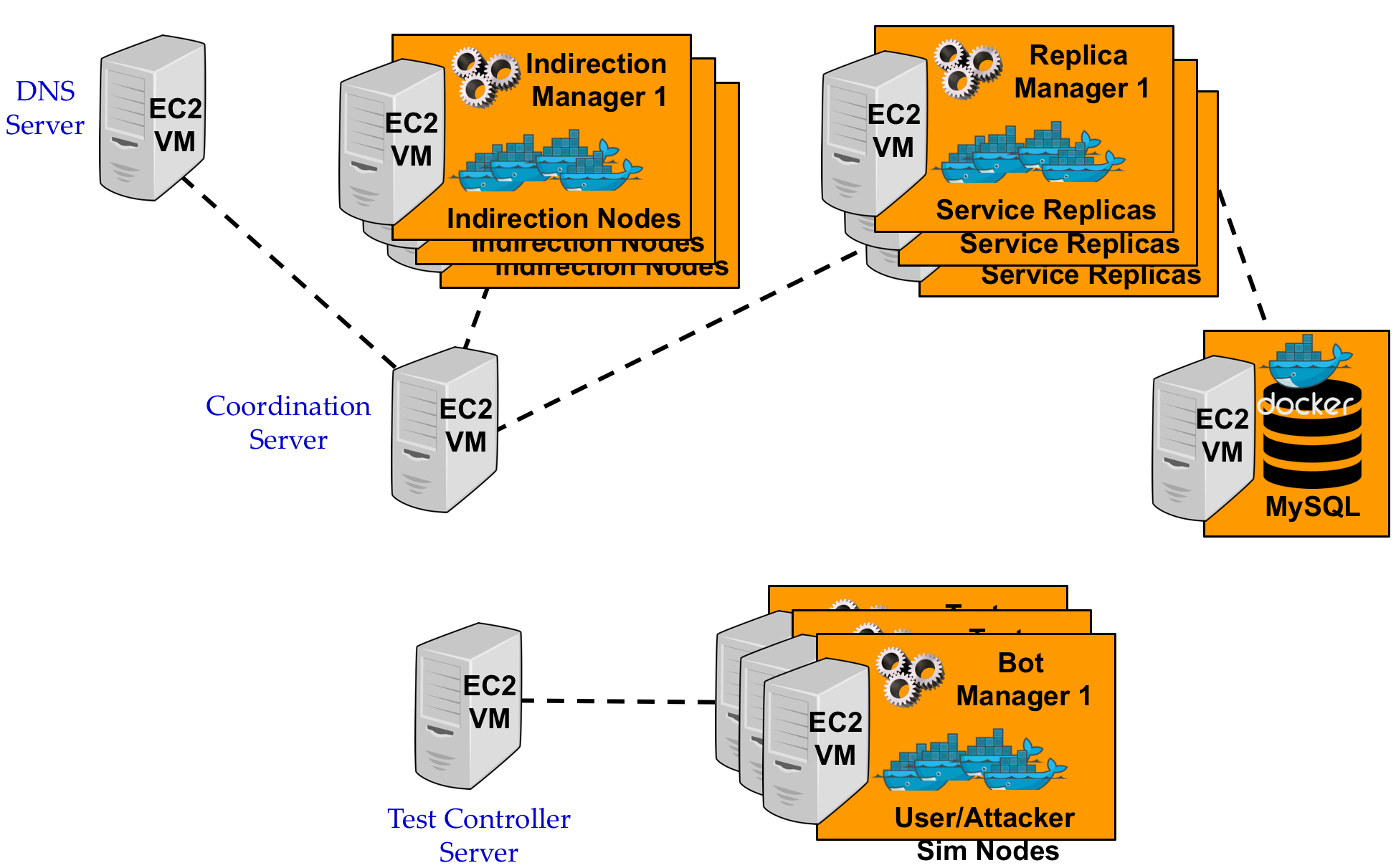}
\caption{Physical architecture of the defense and test setup on AWS.}
\label{fig:physical-arch}
\end{center}
\end{figure}

% In a real deployment, firewall rules will only allow traffic into the system to the publicly accessible components.

\subsection{Test Experimental Setup}

To facilitate testing we have also built test infrastructure using a similar design as the defense. The test infrastructure has a 
test controller VM which configures and launches various attacks and load benchmarks against the defense system. 
The test controller communicates with any number of Bot Manager VMs. Each BotManager VM instantiates multiple 
Docker containers within the VM to launch attacks from multiple IPs and simulate many users accessing the system. The 
test architecture can create both nominal clients and attackers running various types of volumetric attacks (targeting proxies)
and application level attacks (targeting replica application servers).

\subsection{Experimental Capabilities}

Using the described setup the team, has implemented multiple types of 
proactive and reactive defenses:
\begin{itemize}
\item Random shuffling of client assignments 
among attacked servers  (followed by attack detection)
to create ``clean" servers to label nominal clients.
\item Random shuffling of client assignments  
among {\em all} servers  (followed by attack detection)
as a basis for a client reputation system.
\item Fission of attacked servers (followed by attack detection). 
%\cf next section.
\item Random modification of proxy identity (IP address) to 
proactively defeat attacker's reconnaissance phase
 \cite{Stirling-TR}. 
\end{itemize}
Shuffling and fission can also be used proactively for purposes of load balancing to deal with overloaded servers.

In addition, the system is designed to support both stateful and stateless clients. Stateless clients are able to be redirected to
other servers without transferring state, \eg HTTP(S) traffic. 
Stateful clients must transfer state to the new target server, \eg
multimedia streaming or VoIP session establishment. 
Members of our  implementation team are working on efficient 
stateful client reassignment techniques (leveraging prior work
such as \cite{clark2005live,hines2009post}) 
%  \cite{Costas,VMTorrent} 
and porting our experimental framework to other public clouds, \eg
GCE, Azure.

%The following numerical evaluations 
%%in Section \ref{sec:performance} 
%agree with preliminary results using our
%experimental framework on AWS.

\section{Preliminary definitions and binomial model assuming a fixed number of clients per server}\label{sec:def}

In the following, let
\begin{itemize}
\item $M$ be the number of servers
\item $U=\E u$ be the mean number of nominal users 
\item $K=\E \kappa$ be the mean number of attacker users
\end{itemize}

In the following, we will evaluate performance primarily in terms
of the number of nominal (not attacking) clients entirely
unaffected  by or liberated from the attack.
Overhead includes number of additional containers/VMs used and
delays experienced by clients owing to the defense itself.

Assume an equal number of users are assigned to each server.
The following analysis is based on the binomial distribution:
\beqa
a(\cdot) =  {\sf binom}(v,p)~\mbox{where}~v=\frac{U+K}{M},~p=\frac{K}{U+K} \\
i.e.,~ a(i) = \binom{v}{i} p^i(1-p)^{v-i} ~ \mbox{for}~ i\in\{0,1,2,...,v\}.
\eeqa
So, $a(i)$ is the probability that a server has $i$ attacking/heavy-hitter
clients.
Note that $K =\E \kappa = Mvp$ and $U=\E u = Mv(1-p)$. 
Also note that  the standard deviations
$\sigma(\kappa) = \sqrt{Mvp(1-p)}=\sigma(u)$ so that
\beqa
\frac{\sigma(\kappa)}{\E \kappa} =  \sqrt{\frac{U}{K(U+K)}} & \mbox{and} &
\frac{\sigma(u)}{\E u} =  \sqrt{\frac{K}{U(U+K)}}.
\eeqa
So, $u\approx U$ and $\kappa \approx K$ with high probability
when $U$ is large and on the order of $K$, \ie $K\sim U \gg 1$.

The servers are designed to simultaneously handle  the 
{\em typical} load of $v$ nominal users.
Let $A \in\{1,2,...,v\}$ be the number of attacking clients required
to overload the server. So, the probability that a server is overloaded is
taken to be
\beqa
\Omega & := & \sum_{i=A}^v a(i).
\eeqa

In \cite{Stirling-TR}, we consider the case where
different servers can have different numbers of clients,
and so use a type of Stirling distribution of
the second kind to model attacker/heavy-hitter assignments
to servers, instead of a simpler binomial distribution.

\section{Shuffling client-to-server assignments}\label{sec:shuffling}

\subsection{Using Neither Server Load State Nor Client History/Reputations}\label{sec:proactive}

Suppose {\em proactive}
 shuffling client-to-server assignments occurs irrespective
of the loads on the servers and prior history of the clients.
The purpose of this proactive shuffling is to maintain a balance
of load among servers so that no one nominal client
experiences total service disruption
even when a ``weak" DDoS attack (or flash crowd) is occurring.
In this section, the overload state of a server is not checked by the
multiserver system, \ie the system is stateless.

\begin{figure}[h]
\begin{center}
\includegraphics[width=\columnwidth]{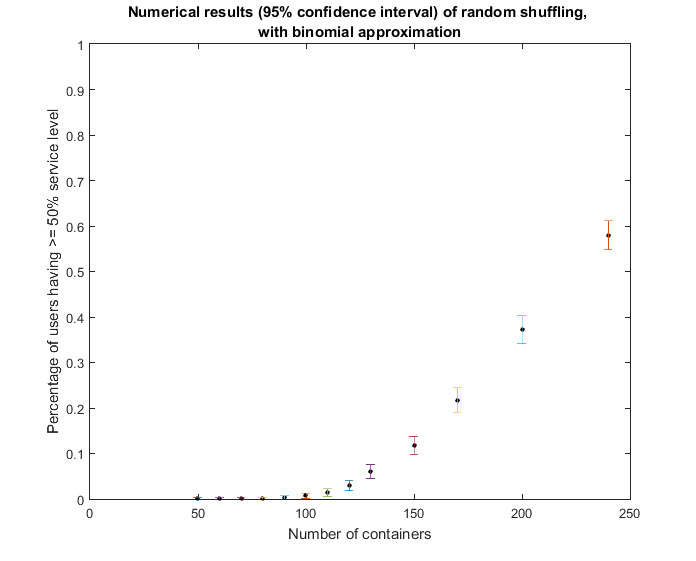}
\caption{(\ref{nominal-client-free}) for 
50\% in-service time of nominal clients over $S=10$ shuffles
as a function of the number of servers $M$, with
$U = 1000$ nominal clients, $K=200$
attackers/heavy-hitters, and $A=1$ attacker/heavy-hitter
can overload a server}\label{fig:nominal-client-free-binom}
\end{center}
\end{figure}

Consider the point-of-view of a given nominal
(not attacking nor heavy-hitter) client. 
The probability that this nominal client is part of an overloaded server is
\be\label{omega-def}
%\omega & := & \left.\sum_{i=A}^{v-1} a(i)\middle/ \sum_{i=0}^{v-1} a(i) \right. \\
\omega & := & \sum_{i=A}^{v-1} \binom{v-1}{i} p^i(1-p)^{v-1-i} 
\ee
That is, $\omega$ is the fraction of time a nominal client is
in an overloaded server.
For a nominal client active 
over a period of $S$ shuffles,  the number of
times it is assigned to a not-overloaded server
(\ie times that the client is ``in service") is
distributed binomially $(S,1-\omega)$ distributed, \ie
\beqa
b(k) = \binom{S}{k} (1-\omega)^k \omega^{S-k} ~~\mbox{for $k=0,...,S$,}
\eeqa
again here  taking fixed $U$ %=S\lambda$ 
and $K$ in the expression for
$\omega$ (\ref{omega-def}).
So the probability that such a nominal client is in-service
$x$\% of the time is
\begin{eqnarray}\label{nominal-client-free}
\sum_{k = \lceil S x/100\rceil}^S  b(k).
\end{eqnarray}
In  Figure  \ref{fig:nominal-client-free-binom},  
the in-service time (\ref{nominal-client-free}) is plotted with 95\% 
confidence bars (2 standard deviations, after $S=10$ shuffles) 
as a function of the number of servers $M$ for  $U=1000$ nominal
clients and $K=1000$ attackers.

\paragraph{Consecutive service interruptions}
Given a sequence of random shuffles of client-to-server assignments,
a given nominal client will experience alternating periods of 
assignment to overloaded servers (basically a period of service outage)
and non-overloaded servers (service restoration).
The period of consecutive shuffles assigned to overloaded servers is 
geometrically
distributed with parameter $\omega$. More specifically, the
probability that a service outage lasts for $s\geq 1$ shuffle-periods\footnote{A
``shuffle period'' is the assumed fixed time between consecutive shuffles.}
\beqa
\omega^s(1-\omega).
\eeqa
The mean and variance of service outage is therefore
$\mu=\omega/(1-\omega)$ and $\sigma^2=\omega/(1-\omega)^2 $, respectively.
Similarly, the period of time where service is restored is
geometrically distributed with parameter $1-\omega$.

For example, a long-lived streaming media session  connection will thus
experience  {\em intermittent availability}. Suppose that during
periods of service restoration, the client-side playout
buffer is primed with $T$ seconds of content. Service
will be only infrequently interrupted from the client's point-of-view if
$T>\mu + 2\sigma$ (mean plus two standard deviations of service outage).

% discussed in  ../fission/overview.tex
%the overhead of migrating a streaming media server? particularly
%the disk IO required to prime the cache in the new VM/container
%as the old VM/container

For another example, consider a SIP server bank under DDoS attack so that
new media sessions cannot be established by some overloaded servers.
%https://en.wikipedia.org/wiki/Session_Initiation_Protocol#/media/File:SIP_call_flow_between_UA,_Redirect_Server,_Proxy_and_UA.png
Under shuffling,  SIP servers will be only intermittently overloaded so
that  call set-up times will be longer on average but calls will
eventually be set up.

\paragraph{Effect of client churn}
Now consider clients with finite lifetimes.
If the mean arrival rate of
nominal clients is $\lambda_{\rm n}$ per shuffle and
the mean lifetime of nominal clients  is $1/\mu_{\rm n}$ shuffles, 
then by Little's formula, the steady-state
mean number of nominal clients is $U=\E u = \lambda_{\rm n} /\mu_{\rm n}$.
Let $V(u,k)$ be the quantity in (\ref{nominal-client-free}) as
as a function of the number of nominal clients $u$ and attacker/heavy-hitter
clients $k$. We can interpret Figure 
\ref{fig:nominal-client-free-binom} as $V(\E u, \E k)=V(U,K)$.

Assume  the arrivals of new clients  are independent Poisson processes 
and the lifetimes are independent exponentials with mean 
$1/\mu_{\rm n}$ or $1/\mu_{\rm a}$ for nominals and 
attackers/heavy-hitters. Thus, the number of attackers $u$ and
$k$ are independent, Poisson distributed 
random variables\footnote{M/M/$\infty$ queues.}: \ie
for all integers $i\geq 0$,
$$p_x(i) = x^i e^{-x}/i!,$$
where $x$ is the mean and the variance:
$x= K = \E k = \lambda_{\rm a}/\mu_{\rm a}$ or 
$x= U = \E u = \lambda_{\rm n}/\mu_{\rm n}$.
So in this case, 
$$\E V(u,k) ~ = ~ \sum_{x=0}^\infty \sum_{y=0}^\infty V(x,y) p_U(x)p_K(y). $$

\begin{figure}[h]
\begin{center}
\includegraphics[width=\columnwidth]{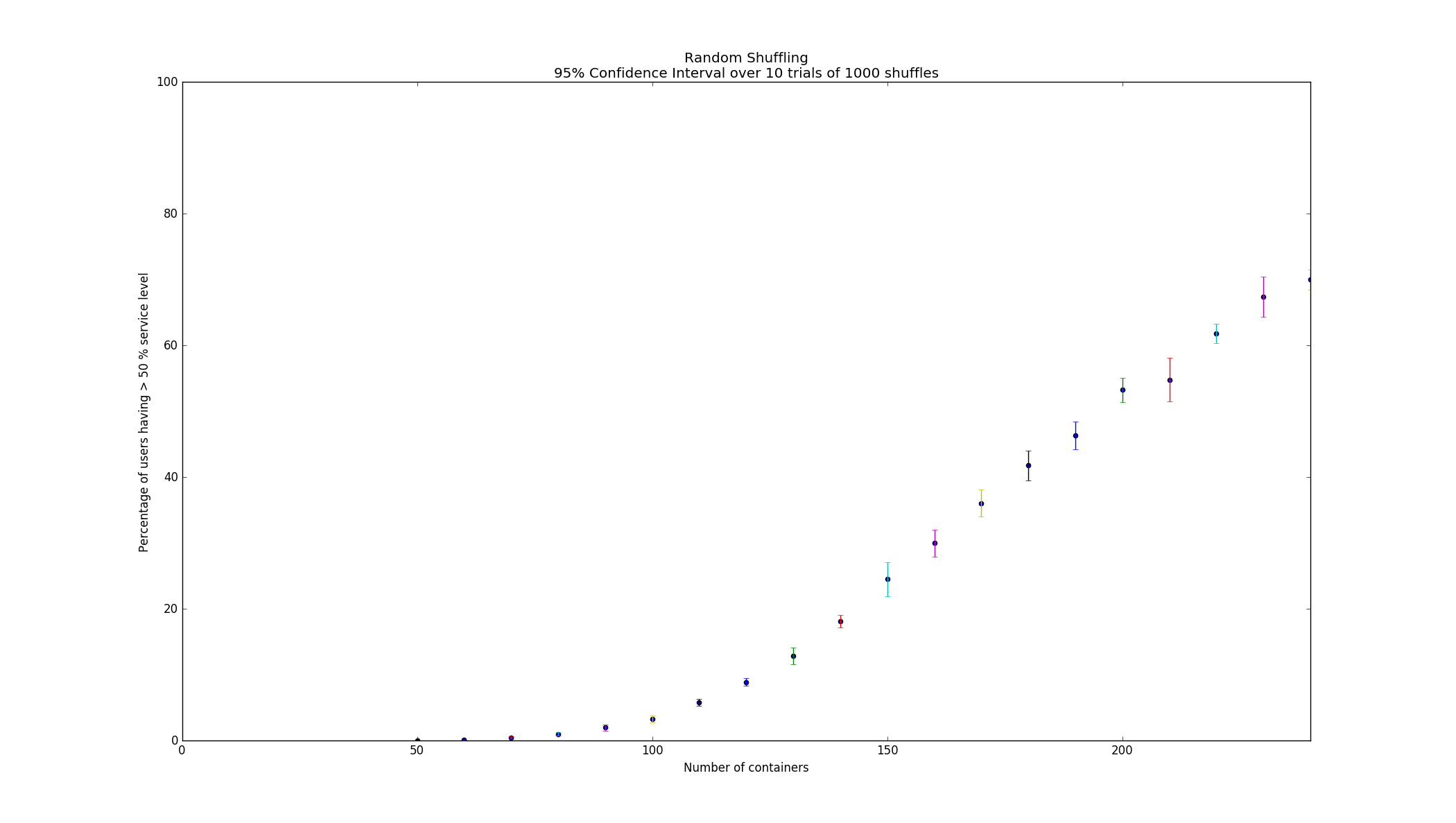}
\caption{ 50\% in-service time of nominal clients 
as a function of the number of servers $M$ under 
client churn}\label{fig:nominal-client-free-sim}
\end{center}
\end{figure}

Naturally, lifetimes and interarrival times are likely not  independent and
exponentially distributed in practice. In Figure
\ref{fig:nominal-client-free-sim}, we plot the
50\% in-service likelihood obtained by simulation for an example where:
 deterministic lifetimes  $1/\mu_{\rm a} = 100$ and $1/\mu_{\rm n} = 10$
shuffles,
arrivals with mean rates $\lambda_{\rm a}=2$ (and standard deviation 2) and
$\lambda_{\rm n}=100$ clients/shuffle (and standard deviation 10)
that are i.i.d. Gaussian processes,
and a single assigned attacker causes server  overload (\ie  $A=1$). 
So, by Little's formula,
$K=2\cdot 100= 200$ and $U=100\cdot 10=1000$ as in  Figure
\ref{fig:nominal-client-free-binom}. Note that
the 50\% in-service likelihood for nominal clients
is higher under this
client churn example than for fixed client populations
given in Figure \ref{fig:nominal-client-free-binom}. 
In other cases, client churn may increase this quantity.

\subsection{Using Server Load State and Client History/Reputations}\label{sec:reputations}

Binomial distributions have in the past been 
used to model reputation systems, \eg \cite{Josang07}.
Consider a client reputation system (\ie memory of client activity maintained)
such that a high reputation implies a 
nominal client while a low reputation implies a heavy hitter or attacker.
The reputation system works with successive
rounds of randomized shuffling of client-to-server
assignments, where the overload state of each
server is assessed initially and after each shuffle. 
So, in each round, the reputation of clients associated
with a deemed overloaded server decreases by 1, otherwise increases by 1.

\begin{figure}[h]
\begin{center}
\includegraphics[width=\columnwidth]{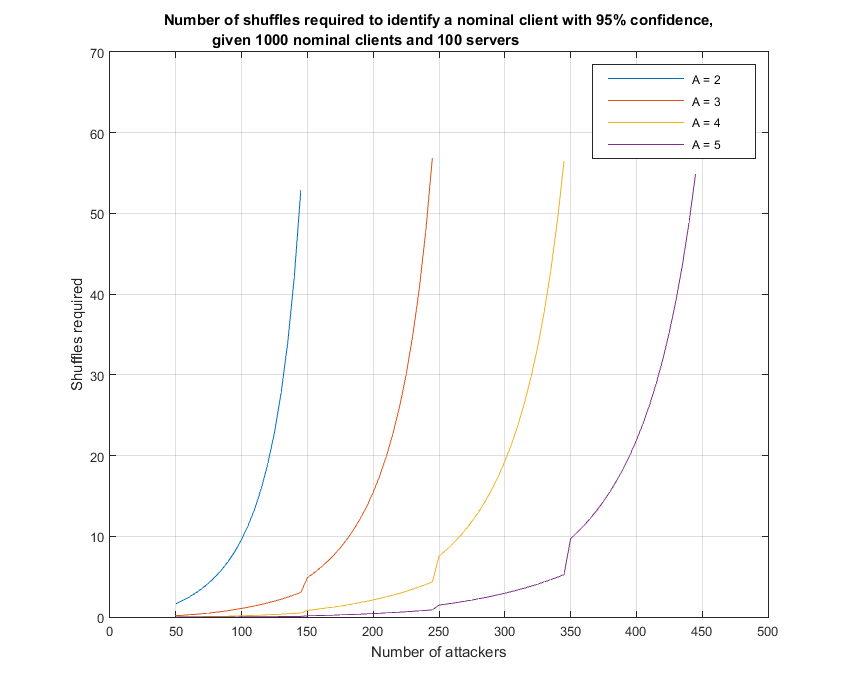}
\caption{Lower bound (\ref{nominal-reput-bound}) on the 
number of shuffles required to deem any client nominal/benign
with 95\% confidence}\label{fig:reputations-nominal}
\end{center}
\end{figure}

Again recall $\omega$ from Equation (\ref{omega-def})
is the probability that a nominal client is part 
of an overloaded server.
After $S$ shuffles, the number of overloaded servers
to which a nominal client is assigned is $\Omega\sim{\sf binom}(S,\omega)$,
with mean $\E\Omega = \omega S$ and variance 
${\sf var}(\Omega)=S\omega(1-\omega)$.
So, the reputation  of a nominal client after $S$ shuffles is
$R_{\rm n} =(S-\Omega)-\Omega= S-2\Omega$ which has mean 
$\E R_{\rm n} = S(1-2\omega)$
and variance 
${\sf var}(R_{\rm n})=4S\omega(1-\omega)$.
So one can deem any client nominal ($R_{\rm n}>0$)
{\em within a 95\% confidence interval} if  $\omega<0.5$ 
and the number of shuffles $S$ is sufficiently large so that,
\be
0~<~\E R_{\rm n} - 2\sqrt{{\sf var}(R_{\rm n})} & = & 
S(1-2\omega) - 2\sqrt{4S\omega(1-\omega)}, \nonumber\\
i.e.,~~ S & > &  \frac{16\omega(1-\omega)}{(1-2\omega)^2},
\label{nominal-reput-bound}
\ee
\ie 2 standard deviations corresponds to 95\% confidence
(for 99\% confidence, use 3 standard deviations).
Figure \ref{fig:reputations-nominal} depicts this lower
bound as a function of the number of attackers,
for $U=1000$ nominal clients and $M=100$ servers.
These plots are consistent with intuition:
it becomes more difficult to confidently 
detect a nominal/benign client 
\begin{itemize}
\item as the number of attackers/heavy-hitters decreases,  or
\item as the number of attackers/heavy-hitters required to 
overload a server ($A$) increases (attackers can ``hide" in 
servers that are not overloaded).
\end{itemize} 

\begin{figure}[h]
\begin{center}
\includegraphics[width=\columnwidth]{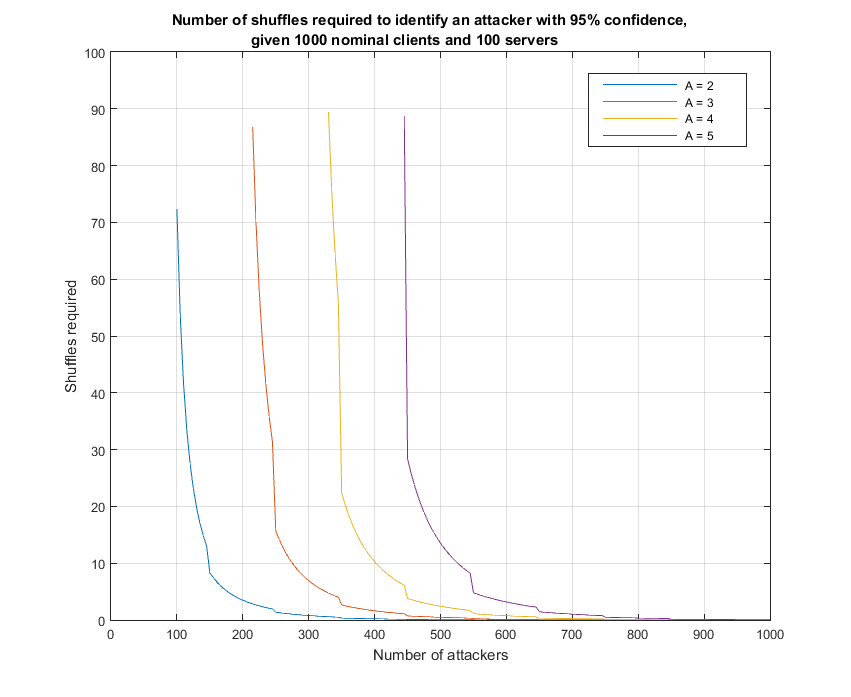}
\caption{Lower bound (\ref{attack-reput-bound}) on the 
number of shuffles required to deem any client an attacker/heavy-hitter
with 95\% confidence}\label{fig:reputations-attack}
\end{center}
\end{figure}

Similarly, let $\beta$ be the probability that a heavy-hitter/attacker
is assigned to a not-overloaded server,
\be\label{alpha-def}
\beta & := & \sum_{i=0}^{A-2} \binom{v-1}{i} p^i(1-p)^{v-1-i} 
\ee
After $S$ shuffles, the number of not-overloaded servers
to which a heavy-hitter/attacker client is assigned is 
$B\sim{\sf binom}(S,\beta)$,
with mean $\E B = \beta S$ and variance 
${\sf var}(B)=S\beta(1-\beta)$.
So, the reputation  of an heavy-hitter/attacker
 client after $S$ shuffles is
$R_{\rm a} =B - (S-B)= 2B-S$ which has mean 
$\E R_{\rm a} = S(2\beta-1)$
and variance 
${\sf var}(R_{\rm a})=4S\beta(1-\beta)$.
So one can deem a client an attacker ($R_{\rm a}<0$)
within a 95\% confidence interval if  $\beta<0.5$ 
and the number of shuffles $S$ is sufficiently large so that,
\be
0~>~\E R_{\rm a} + 2\sqrt{{\sf var}(R_{\rm a})} & = & 
S(2\beta-1) + 2\sqrt{4S\beta(1-\beta)}, \nonumber \\
i.e.,~~ S & > &  \frac{16\beta(1-\beta)}{(1-2\beta)^2}.
\label{attack-reput-bound}
\ee
Figure \ref{fig:reputations-attack} depicts this lower
bound as a function of the number of attackers,
for $U=1000$ nominal clients and $M=100$ servers.
These plots are also consistent with intuition as 
Figure \ref{fig:reputations-nominal}.

%Alternatively, 
%the defense may wish simply to find the detection reputation-threshold
%above which the client is {\em more likely} nominal and below which
%the client is {\em more likely} an attacker or heavy hitter.
%This threshold is simply the point where the two 
%distributions of $R_{\rm a}$ and $R_{\rm n}$
%discussed above meet (\ie close to zero, 
%between $\E R_{\rm a}$ and $\E R_{\rm n}$).

\begin{figure}[h]
\begin{center}
\includegraphics[width=\columnwidth]{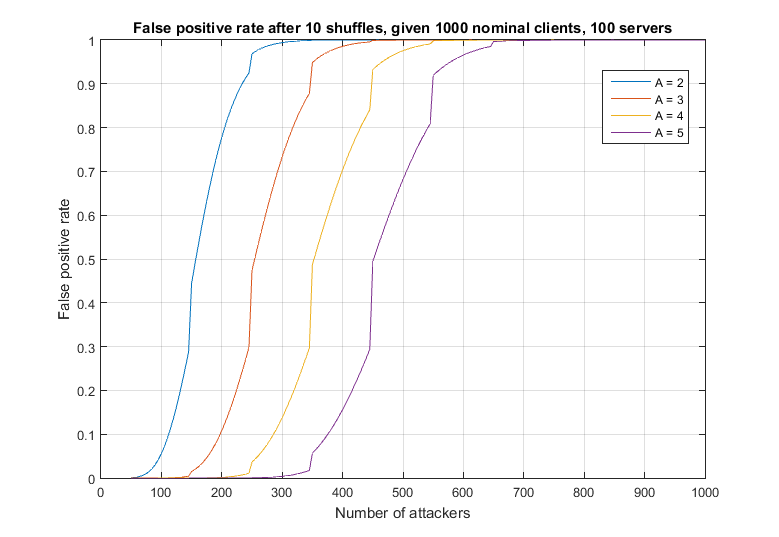}
\caption{False positive rates (\ref{false_positive_rate}) for the example of
Figure \ref{fig:reputations-attack} 
after 10 shuffles}\label{fig:reputations-attack-fp}
\end{center}
\end{figure}

\begin{figure}[h]
\begin{center}
\includegraphics[width=\columnwidth]{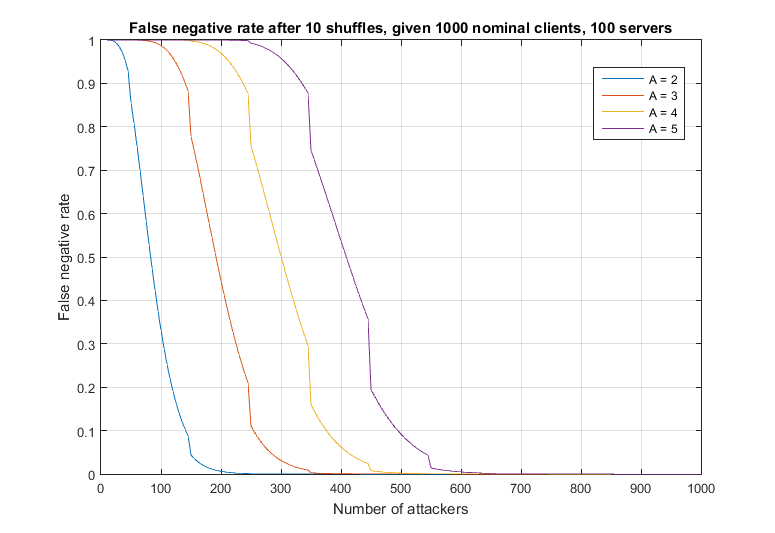}
\caption{False negative rates (\ref{false_negative_rate}) for the example of
Figure \ref{fig:reputations-nominal} 
after 10 shuffles}\label{fig:reputations-nominal-fn}
\end{center}
\end{figure}

If the aim is to detect attacker/heavy-hitter clients with high confidence
using (\ref{attack-reput-bound}), then 
one should also be interested in the
probability that a nominal client is deemed an attacker/heavy-hitter,
\ie the {\em false positive} rate (otherwise, simply take {\em all} clients
as nominal to ensure that all truly attacker/heavy-hitter clients are
detected).   One typically wants to 
detect attacks with high confidence subject to a bound $\phi$ on
false positives. The bound on false positives may be higher or
lower depending on the desired tradeoff between usability (low false
positives) and security (high true positives).
That is, one can take the {\em maximum} of the lower bound
on the number of shuffles required to (i) detect attackers
with sufficiently high confidence (\ref{attack-reput-bound}) 
and (ii) 
limit the the false positive rate to $\phi$, the latter
given by
\be
\phi & \geq & \P(R_{\rm n} < 0) \\
   % & = &   \P(S-2\Omega < 0) \\
    & = &   \P(\Omega > S/2) \\
  & = &   \sum_{k=\lceil S/2\rceil}^S \binom{S}{k}\omega^k(1-\omega)^{S-k}.
\label{false_positive_rate}
\ee
For the example from Figure
\ref{fig:reputations-attack}, the
 false positive rates are depicted in  Figure
\ref{fig:reputations-attack-fp}.
%Indeed, in some cases, the aim is to maximize the
%confidence in detecting attacker/heavy-hitter clients subject to 
%an upper bound on the false positive rate.
Similarly, when trying to detect nominal clients with high confidence 
using (\ref{nominal-reput-bound}), 
one should also consider the {\em false negative} rate  
\be
\P(R_{\rm a} > 0) 
  % =   \P(2B-S>0)
   =   \P(B>S/2) 
 =   \sum_{k=\lceil S/2\rceil}^S \binom{S}{k}\beta^k(1-\beta)^{S-k}.
\label{false_negative_rate}
\ee
For the example from Figure
\ref{fig:reputations-nominal}, the
 false positive rates are depicted in  Figure
\ref{fig:reputations-nominal-fn}.
A more extensive study of reputation systems in this context is
conducted by our collaborators in \cite{reputs-columbia}.
%https://drive.google.com/file/d/0BxGYotl5qUnNUTM5WXZOcDZCa2c/view?usp=sharing

\paragraph{Future Work}
 In future work,
we will consider reputation systems of clients with finite lifetimes.
Considering churn among clients motivates an autoregressive
and/or moving average approach
to reputations wherein older history of client involvement
in overloaded servers has less impact on their current reputations.
For example, if a client is assigned to a server subsequently
deemed overloaded, then their reputation changes according
to $R\rightarrow \alpha R - (1-\alpha)$, otherwise $R\rightarrow \alpha R  
+ (1-\alpha)$. So, the expected reputation of a long-lived
heavy-hitter/attacker  client capable of  overloading a server
by itself is $-1$.

\begin{comment}
Suppose even after DDoS attack detection,
\underline{all} servers are shuffled even those
not detected as under attack, \ie \underline{initially} the aim is
not quarantine of the attackers.
After each shuffle, attack detection is performed on each server.
After each attack detection, the reputation of all clients
in attacked servers is reduced relative to those in not-attacked servers.
After a certain number of shuffle/detection steps, 
all clients with sufficiently low reputation are quarantined.
Note  that detection may not be perfect,  so 
may want to quarantine not just
the clients with {\em lowest} reputation. An obvious trade-off
between false positives and missed detections ensues when
choosing the threshold that deems a client an attacker.
\end{comment}

% the following included in ../fission/overview.tex
\begin{comment}
In the case of reacting to an intense  DDoS attack,
if all servers are initially under overloaded (attacked) then one can
\begin{itemize}
\item sequentially sequester/queue a fraction of all clients to an idle server
until not all servers are overloaded,
\item employ shuffling/detection with reputation to quarantine some of the
remaining clients, 
\item introduce some of the sequestered clients into service and repeat.
\end{itemize}
\end{comment}

%Might instead assign a single reputation to clients with common address prefixes (as reputations are commercially used for email spam)

%\input{../reputations/reputs}

\subsection{Using Server Load State but not Client History/Reputations}\label{sec:load-no-reputs}

Now suppose that the system does determine the overload state of
each server, recall Section \ref{sec:overhead}.
Again, a DDoS attack on a multiserver system may be detected when a
sufficient number $S$ of servers are determined to be overloaded.
In the absence of such a DDoS attack detection, the system
sequentially shuffles client-to-server assignments as described
in Section \ref{sec:proactive}.

On the other hand, if DDoS attack is detected, attacked
servers may perform local fission  operations to 
identify and quarantine attacking clients, recall
Section \ref{sec:fission0} and see \cite{fission-TR}.

\section{Quarantine Shuffling Based on Server Load State}\label{sec:shuffling-quarantine}

Here suppose that a DDoS attack has been detected because
a sufficiently large number $S$ of servers are overloaded.
Under quarantine shuffling, only servers deemed overloaded (under attack)
will shuffle client-to-server assignments among themselves.
By performing sequential such shuffle-detections, the number of overloaded
servers is nonincreasing, \ie  periodically and by chance, 
some formerly overloaded servers are assigned only nominal clients
and those clients are then ``liberated" from attack.
Again, the objective here is quarantine not load balancing.

\subsection{Without Using Client Histories (Reputations)}

Let $L_0$ be the expected number of clients which are not in the attacked servers, which is a function of $K,U,M$:
$$L_0(K,U,M) = Ma(0)v.$$

The $(U+K)Q_1/M$ users including $K$ attackers are randomly reassigned (shuffled
) to the $Q_1$ initially attacked servers, 
where $Q_1\sim {\sf binom}(M,1-a(0))$.

The number of freed clients after one shuffling step is
$$
L_0(K,U,M) + \sum_{n=Mp}^{M\wedge K} B(n) L_0(K,vn-K,n),
$$
This is easily extended to plural shuffle-detection stages
as \cite{Stirling-TR}.
To generalize  to the case where the (proxy) server can tolerate up to 
$A\geq 0$ attackers ($A\leq v$), instead use
$Q_1\sim {\sf binom}(M,1-\sum_{k=0}^A a(k))$.
Moreover, one can initially mobilize only $M-H$ servers
and employ $H$ hot-spare (standby) servers for shuffling only after
a DDoS attack has been detected. One can easily
compare quarantine performance for difference values
of $H$ for the binomial model again as we have for
the Stirling model in \cite{Stirling-TR}.

%\subsubsection{Simulation and numerical results - shuffling}

\begin{figure}[h]
\begin{center}
\includegraphics[width=\columnwidth]{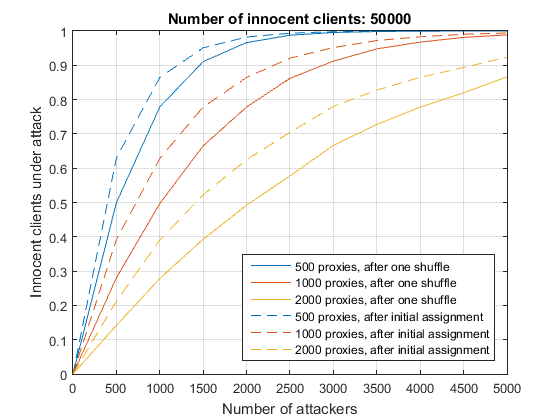}
\caption{Mean fraction of affected nominal/innocent clients
after single shuffle stage as a function of the number of
proxies (or replica application servers) - \ie difference between 
dotted and solid lines of the same color}\label{fig:shuffling-gain}
\end{center}
\end{figure}

Figure \ref{fig:shuffling-gain} shows a typical result of our shuffle-quarantine
study wherein a maximum gain of about 20-25\% liberated nominal clients
occurs when the number attackers is about equal to the
number of servers, \ie $K\approx M$. This is consistent with
the results of \cite{Stirling-TR} using a Stirling instead of
a binomial distribution (again, the former allowing for different numbers
of clients per server and only considers assignments of
attackers/heavy hitters). During a DDoS attack or flash crowd,
servers can be dynamically spun up, based on the performance
of previous shuffles, and added to the pool of overloaded
servers for purposes of shuffling client assignments,
recall Section \ref{sec:existing-defenses}.

\subsection{Future Work: Using Client Histories}

% Recall reputation systems discussed above...

In future work we will explore how a
server may have a range of load state (not just
overloaded or not overload) and determining the
overload state of a server may be prone to error.
The degree of overload of a server can be reflected in different
reputation increments of its current clients. 

In future work we will also explore how shuffling 
might not be purely random but could instead tend to cluster clients
with lower reputations,
\ie white-listing clients with high reputations,
black-listing (quarantine) of clients with low reputations.
That is, clients with sufficiently low reputations can be quarantined.
We will also consider the impact of client churn 
on quarantine performance.

\section{Summary and Outlook}\label{sec:summary}

In summary, for a generic proxied multiserver system,
we have studied proactive (load balancing before 
attack/flash-crowd detection) and reactive (after DDoS attack detection)
techniques  of shuffling
client-to-server assignments.
We studied cases where client-to-server assignment shuffling was based on:
\begin{itemize}
\item not using server load state in Section  \ref{sec:proactive}
considering service outages of nominal clients;
\item server load state  (assessed between shuffles) and
history of client involvement in overloaded servers
(client reputations) in Section \ref{sec:reputations} considering
the number of shuffles required to detect nominal
or attacker/heavy-hitter clients with high confidence,
and corresponding false negative or false positive rates; and
\item server load  for attacker/heavy-hitter quarantine  in
Section \ref{sec:shuffling-quarantine} where we found optimal
performance when the number of attackers was approximately
the number of targeted servers.
\end{itemize}
The given numerical results are typical of a more extensive and
consistent 
numerical and simulation study.
Our numerical and simulation studies will inform
a nascent  emulation platform for DDoS attack/defense
experimentation on public-cloud computing platforms currently
being developed on AWS (and in the future, on GCE and
Azure).

We also discussed how to use  server load state without 
quarantine in Section \ref{sec:load-no-reputs} and
discussed future work on, \eg the use of client reputations with
quarantine, different degrees of server load, and client churn.
%We are also working on implementations of 
%moving-target defenses against reconnaissance and fission based defenses.
%We are also working on promising 
%techniques of fission of overloaded servers
%\cite{fission-TR} and on 
%a moving-target defense wherein proxy servers periodically
%change their IP addresses to defeat botnet reconnaissance,
%\eg \cite{Stirling-TR}.

%\subsubsection*{Acknowledgments}
%This research was supported by
%DARPA XD3 grant, gifts  of Amazon AWS and Google GCE credits,
%and a Cisco Systems URP gift.

\bibliographystyle{abbrv}
\bibliography{../../latex/stirling,../../latex/ddos,../../latex/cloud,../../latex/p2p,../../latex/unstruct,../../latex/stochastic,../../latex/IoT,../../latex/ref_cheng}

\section*{Appendix: Modeling of Proactive Moving-Target Defense to Defeat Botnet
Reconnaissance}

\begin{figure}[h]
\begin{center}
\includegraphics[width=\columnwidth]{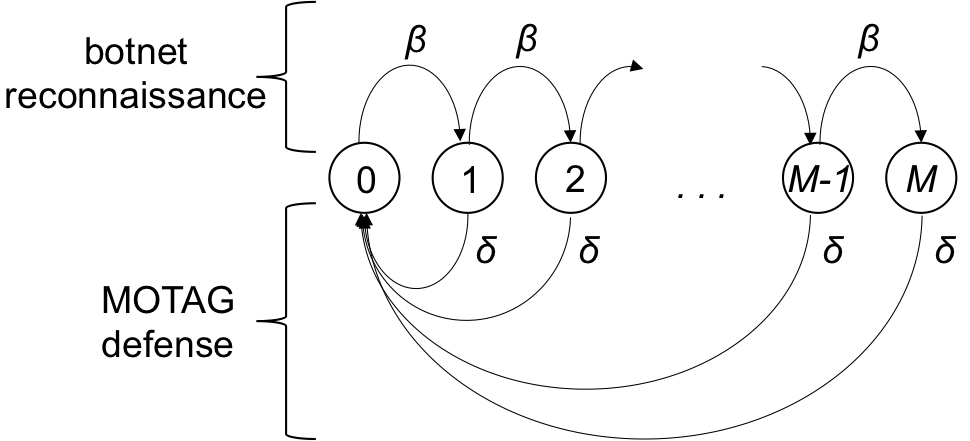}
\caption{Transition rate diagram for a Markov model of the number of
proxies known to the botnet $V$
when $q_{n,n+1}=\beta$}\label{fig:TRDbeta}
\end{center}
\end{figure}

We describe a simple Markov model; see \cite{Stirling-TR} for another
approach.
Let $V_t$ be the number of proxy server identities (IP addresses)
known to the botnet at time $t$. Assume there
are $M$ proxies. We can model $V$ as a continuous-time
Markov chain with transition rates to/from states $n$ of V are
\begin{eqnarray*}
q_{n,n+1} & = & \beta (M-n) ~~\mbox{or just}~~ \beta~~~~\mbox{for}~
0\leq n\leq M-1\\
q_{n,0} & = & \delta ~~~~~~~\mbox{for}~1\leq n\leq M
\end{eqnarray*}
where parameters
\begin{itemize}
\item  $\beta$ captures how quickly the botnet can ascertain the identity
of a proxy (including {\em e.g.}, DNS transaction delays),  
which may become more difficult with fewer 
proxies unknown to the botnet ($M-V$).
%, where  $\beta$ 
%could be taken proportional to the probing rate of the botnet
%(inversely proportional to mean successful probing time, $T_{\rm pr}$); and
\item $\delta$ captures how frequently the defense changes proxy
identities.
% {\bf and the assumption that bots cannot handle changes in their proxy's IP address}.
%$\delta$  could be taken inversely proportional to the 
%mean ``shuffle time" of the defense, $T_{\rm sh}$.
\end{itemize}
See Figure \ref{fig:TRDbeta} for the case $q_{n,n+1}=\beta$.
The invariant distribution $\pi$ of (not time-reversible) $V$  satisfies 
the balance equations $\pi^{\rm T}Q=0^{\rm T}$, where the transition rate
matrix $Q=[q_{n,m}]$ and $q_{n,n}= - \sum_{m\not = n} q_{n,m}$.

\begin{figure}[h]
\begin{center}
\includegraphics[width=\columnwidth]{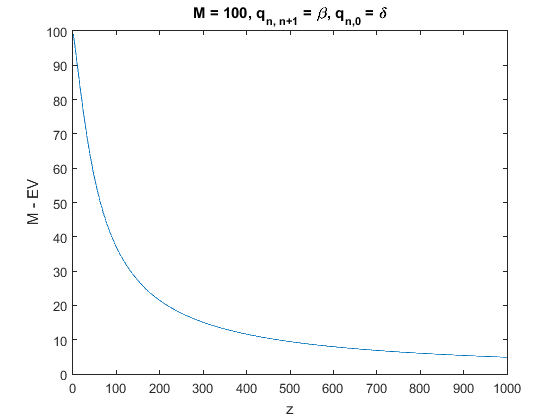}
\caption{Steady-state mean number of proxies unknown to the 
botnet $M-V$ when $q_{n,n+1}=\beta$}\label{fig:beta}
\end{center}
\end{figure}

%That is, the (linearly dependent) balance equations are
%\begin{eqnarray*}
%-\pi_0 \beta + \delta\sum_{i=1}^M \pi_i & = & 0\\
%\pi_{i-1} \beta - (\beta+\delta)\pi_i & = & 0,~~0<i<M\\
%\pi_{M-1}\beta - \pi_M \delta & = & 0
%\end{eqnarray*}
%So 
Let
$$z=\frac{\beta}{\delta} %=\frac{T_{\rm sh}}{T_{\rm pr}}
~\mbox{and}~ x = \frac{\beta}{\beta+\delta} = \frac{z}{z+1}.$$ 
If $q_{n,n+1} = \beta$,  then
\begin{eqnarray*}
%\pi_M  
%& = & %z\pi_{M-1}=
%zx^{M-1}\pi_0\\
\pi_i & = & %x \pi_{i-1} = 
x^i  
(1-x)
, ~~ 0\leq i\leq M
%, ~~ 0<i<M
%\pi_0
\end{eqnarray*}
%Since $\sum_{i=0}^M \pi_i = 1$,
%\begin{eqnarray*}
%\pi_0 & = &  
%\left(1+\sum_{i=1}^{M-1} x^i +zx^{M-1}\right)^{-1} 
%~ = ~  1-x 
%\end{eqnarray*}
Thus, we have in steady state that,
\begin{eqnarray*}
{\sf E} V & = & \sum_{i=0}^M i \pi_i  \\
 & = & \frac{x(1-x^M)}{1-x} ~=~ z\left(1-\left(\frac{z}{z+1}\right)^M\right)
\end{eqnarray*}
and
$$ M~ \geq~ {\sf E} V~ \uparrow ~  M~ \mbox{as $z\rightarrow \infty$}.$$
See Figure \ref{fig:beta}.

\begin{figure}[h]
\begin{center}
\includegraphics[width=\columnwidth]{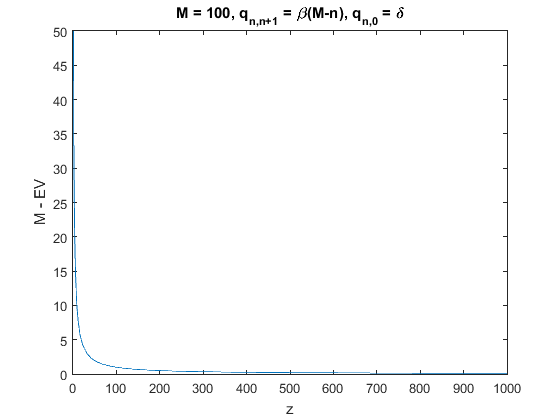}
\caption{Steady-state mean number of proxies unknown to the 
botnet $M-V$ when $q_{n,n+1}=\beta(M-n)$}\label{fig:betaM-n}
\end{center}
\end{figure}

If we suppose that the DNS lookup is randomized instead of round robin,
then the probing time to find an unknown  proxy may be a decreasing
function of the number of remaining unknown proxies, $M-V$.
See Figure \ref{fig:betaM-n} for the case where $q_{n,n+1} = \beta(M-n)$.
There are other, similar illustrative
models that can be explored, \eg  by taking
$q_{n,n+1} = \beta(M-n)/M$ instead.

\begin{comment}
\subsection*{Alternative model:}
Let $A$ be distributed as the time required to modify the IP address of a proxy, with mean $T_{\rm Sh}:=\E A$.
Let $D$ be distributed as the time required to disseminate a proxy's IP address in the botnet, with $T_{\rm Pr} := \E D$.
Assume all proxies have back-to-back sessions with some probing bot.

Arguing as for Little's formula,
the average fraction of fresh proxy IP addresses in the botnet is
\beqa
\frac{\E(A-D)^+ }{\E A} \cdot
\frac{\E V}{M} & = &
\frac{\E(A-D)^+ }{\E A} (1-(1-\frac{1}{M})^K)
\eeqa
If $A$ and $D$ are modeled as exponentially distributed, then
\beqa
\E(A-D)^+ /\E A & =& 1- (1+T_{\rm Sh}/T_{\rm Pr})^{-1}.
\eeqa
For the simple exponential model,
this quantity is plotted in Figure \ref{fig:motag} as a 
(convex) function of
$T_{\rm Sh}/T_{\rm Pr}$. Note that the function is convex,
and half of the servers are not known to the botnet when
$T_{\rm Sh}=T_{\rm Pr}$, but to achieve 75\% not known 
requires $T_{\rm Sh}=3T_{\rm Pr}$.

\begin{figure}[h]
\begin{center}
\includegraphics[width=\columnwidth]{../slides/motag.png}
\caption{Performance of moving-target defense against botnet
reconnaissance phase.}\label{fig:motag}
%curve corresponds to $T_{\rm Sh} = T_{\rm Pr}$, red area is ...}\label{fig:motag}
\end{center}
\end{figure}

\end{comment}
\end{document}